\documentclass{nrc1}             
\usepackage{graphicx}
\usepackage{epsfig,epsf,graphics,psfig,lscape,url}
\usepackage[square, comma, sort&compress]{natbib}
\usepackage[french,english]{babel}

\journal{Can.\ J.\ Phys.}
\volyear{XX}{2010}
\begin{document}
\setlength{\arraycolsep}{2.5pt}             
%
%
%
%
\def\newblock{\hskip .11em plus .33em minus .07em}
%
%
%
%
%
%
\title{New Atomic Data for Trans-Iron Elements and Their Application to Abundance Determinations in Planetary Nebulae}

\author{N.\ C.\ Sterling}
\address{Department of Physics and Astronomy, Michigan State University, 3248 Biomedical Physical Sciences, East Lansing, MI 48824-2320, USA}\correspond{sterling@pa.msu.edu}\IDnote{NSF Astronomy and Astrophysics Postdoctoral Fellow}
\author{M.\ C.\ Witthoeft}
\address{NASA Goddard Space Flight Center, Code 662, Greenbelt, MD 20771, USA}
\author{D.\ A.\ Esteves}
\address[unr]{Department of Physics, MS 220, University of Nevada, Reno, NV 89557, USA}
\author{R.\ C.\ Bilodeau}
\address{Department of Physics, Western Michigan University, Kalamazoo, MI 49008, USA}\IDnote{Advanced Light Source, Lawrence Berkeley National Laboratory, Berkeley, CA 94720, USA}
\author{A.\ L.\ D.\ Kilcoyne}
\address[als]{Advanced Light Source, Lawrence Berkeley National Laboratory, Berkeley, CA 94720, USA}
\author{E.\ C.\ Red}
\address[als]
\author{R.\ A.\ Phaneuf}
\address[unr]
\author{G.\ Alna'Washi}
\address[unr]
\author{A.\ Aguilar}
\address[als]

\shortauthor{Sterling et al.}
\maketitle

\begin{abstract}

Investigations of neutron(\emph{n})-capture element nucleosynthesis and chemical evolution have largely been based on stellar spectroscopy.  However, the recent detection of these elements in several planetary nebulae (PNe) indicates that nebular spectroscopy is a promising new tool for such studies.  In PNe, \emph{n}-capture element abundance determinations reveal details of \emph{s}-process nucleosynthesis and convective mixing in evolved low-mass stars, as well as the chemical evolution of elements that cannot be detected in stellar spectra.  Only one or two ions of a given trans-iron element can typically be detected in individual nebulae.  Elemental abundance determinations thus require corrections for the abundances of unobserved ions.  Such corrections rely on the availability of atomic data for processes that control the ionization equilibrium of nebulae (e.g., photoionization cross sections and rate coefficients for various recombination processes).  Until recently, these data were unknown for virtually all \emph{n}-capture element ions.  For the first five ions of Se, Kr, and Xe --- the three most widely detected \emph{n}-capture elements in PNe --- we are calculating photoionization cross sections and radiative and dielectronic recombination rate coefficients using the multi-configuration Breit-Pauli atomic structure code AUTOSTRUCTURE.  Charge transfer rate coefficients are being determined with a multichannel Landau-Zener code.  To calibrate these calculations, we have measured absolute photoionization cross sections of Se and Xe ions at the Advanced Light Source synchrotron radiation facility.  These atomic data can be incorporated into photoionization codes, which we will use to derive ionization corrections (hence abundances) for Se, Kr, and Xe in ionized nebulae.  Using Monte Carlo simulations, we will investigate the effects of atomic data uncertainties on the derived abundances, illuminating the systems and atomic processes that require further analysis.  These results are critical for honing nebular spectroscopy into a more effective tool for investigating the production and chemical evolution of trans-iron elements in the Universe.

\PACS{32.80.Fb, 33.60.+q, 34.80.Lx, 34.70.+e, 95.30.Dr, 95.30.Ky, 97.10.Cv, 97.10.Tk, 98.38.Bn, 98.38.Ly, 98.38.Hv}

\end{abstract}

\section{Introduction}
\subsection{Astrophysical Motivation}

The formation of neutron(\emph{n})-capture elements (atomic number $Z>30$) is not as well understood as that of lighter species, and observational abundance determinations are critical for testing and improving theories for the nucleosynthesis of these elements.  Trans-iron elements are predominantly formed via rapid or slow \emph{n}-capture nucleosynthesis (the ``\emph{r}-process'' and ``\emph{s}-process,'' respectively), in which the distinction is based on the relative timescales of \emph{n} captures and $\beta$ decays.  Each mechanism is responsible for the production of roughly half of the \emph{n}-capture element isotopes in the Universe \cite{busso99, sneden08}.

Trans-iron elements are sensitive probes of the nucleosynthetic histories of astrophysical objects and the chemical evolution of the Universe, in spite of their low abundances \cite{sneden08}.  Much of our knowledge of the synthesis of these elements is based on the interpretation of abundances derived from stellar spectra \cite{wally97, busso99, sneden08}.  However, stellar spectroscopy reveals only part of the story of how \emph{n}-capture elements are produced, due to the limited number of elements accessible in stellar spectra, and the difficulty of observing stellar photospheres during certain stages of evolution.  In recent years, nebular spectroscopy has been shown to be a potentially effective tool for investigations of \emph{n}-capture element nucleosynthesis, based on the detection of these species in a large number of planetary nebulae (PNe\footnote{Following astrophysical nomenclature, we abbreviate the singular form planetary nebula as ``PN'' and the plural form as ``PNe.''}) \cite{pb94, sharpee07, sterling08, sterling09}.

The progenitor stars of PNe (1--8 M$_{\odot}$) may experience \emph{s}-process nucleosynthesis during the late asymptotic giant branch (AGB) phase of evolution, which precedes the formative stages of PNe.  In this process, called the ``main'' \emph{s}-process to differentiate it from the ``weak'' \emph{s}-process that operates in massive stars \cite{prantzos90, the07}, free neutrons are produced primarily by the reaction $^{13}$C($\alpha,n$)$^{16}$O in the ``intershell'' region between the H- and He-burning shells.  Fe-peak nuclei experience a series of \emph{n} captures and $\beta$-decays to transform into heavier elements.  In more massive AGB stars ($>4$--5 M$_{\odot}$), the intershell region may reach temperatures  allowing the $^{22}$Ne($\alpha,n$)$^{25}$Mg reaction to efficiently produce neutrons.  The intershell material, enriched in \emph{s}-process nuclei and carbon, can be convectively dredged up to the stellar atmosphere and ejected into the interstellar medium by stellar winds and PN ejection \cite{busso99,herwig05}.

Nebular spectroscopy provides unique information regarding the nucleosynthesis of \emph{n}-capture elements, opening a window to elements that are not detectable in cool giant stars.  For example, nebular spectroscopy provides access to Ge, Se, Br, Kr, Rb, Te, I, Xe, Ba, and Pb, each of which has been identified in the spectrum of at least one PN \cite{pb94, sterling02, sterling03, zhang05, sharpee07, sterling07, sterling08, sterling09, otsuka10}.  With the exceptions of Rb, Ba, and Pb, these elements are not detectable in AGB stars, and in fact had not been detected in any of their sites of origin until their identification in PN spectra.  Therefore little is known about the origins and chemical evolution of these elements beyond theoretical predictions that have not been tested against observational measurements.\footnote{However, recent models of AGB nucleosynthesis have been applied to Se and Kr abundances in PNe \cite{pignatari08, karakas09, karakas10}.}  Moreover, nebular spectroscopy enables \emph{n}-capture nucleosynthesis to be investigated in classes of stars and stages of evolution during which the stellar photosphere is not easily studied.  Intermediate-mass ($\sim$4--8~M$_{\odot}$) AGB and post-AGB stars experience high mass-loss rates that enshroud their atmospheres in optically thick, dusty circumstellar envelopes \cite{habing96, vw03}.  As a consequence, little is known about their contribution to the Galactic inventory of \emph{n}-capture elements.  However, these objects are observable as Type~I PNe (characterized by large N and He enrichments as expected for intermediate-mass stars \cite{peimbert78, stang06}), in which \emph{n}-capture elements have been detected \cite{sterling08, karakas09}.  Similarly, once an AGB star becomes carbon-rich (C/O $>$ 1) the large increase in opacity leads to heavy mass loss \cite{marigo02, cristallo09}, obscuring the photosphere with circumstellar material.  Convective dredge-up during the final thermal pulses present computational difficulties for AGB star models, but can have an important effect on \emph{n}-capture element yields \cite{karakas09}.  The ionized portion of PNe are composed of material enriched by the final thermal pulses during the AGB, and hence provide an excellent venue for constraining the dredge-up efficiency at low envelope mass.  Finally, \emph{n}-capture elements are potentially excellent tracers of C enrichments in PNe, as both are formed in the intershell region of AGB stars and are convectively dredged up simultaneously.  Given the difficulty in accurately determining C abundances in ionized nebulae \cite{kaler83, rs94}, the \emph{s}-process enrichments of trans-iron elements in PNe provide sensitive probes for enrichments of this astrophysically and biologically important element.

\subsection{Atomic Data Needs}

Before the full potential of nebular spectroscopy of trans-iron elements can be realized, several important atomic data problems must be addressed so that \emph{n}-capture element abundances can be accurately determined in ionized nebulae.  One must first determine the abundances of the observed ions, and then correct for those of unobserved ions (wherein the greatest uncertainties lie).  Much of the atomic data needed for these determinations are unknown, due to the recent nature of this field (\emph{n}-capture elements were not identified in any ionized nebula until 1994 \cite{pb94}, and only in the last few years have they been detected in a large number of nebulae -- totaling about 100 objects to date \cite{sharpee07, sterling08}).

For ionic abundances, the atomic data needed include wavelengths, transition probabilities, and effective collision strengths.  For most of the \emph{n}-capture element ions detected in nebulae, energy levels are satisfactorily known, though the NIST compilations \cite{NIST} are less complete than for lighter elements.  One notable exception is Se$^{2+}$, which can have significant fractional abundances particularly in low-ionization nebulae.  The fine-structure energies of the Se$^{2+}$ $^3$P ground term are uncertain by 6.3~cm$^{-1}$ \cite{moore52}, which lead to a wavelength uncertainty of $\sim$10~\AA\ for the transition [Se~III]~$^1D_2$--$^3P_1$~8854.2.  Notably, this line was reported to be detected in the PN NGC~7027 \cite{pb94}, and Sharpee et al.\ \cite{sharpee07} noted that the nearby He~I~8854.14 line showed marked excess relative to other He~I lines in the same series in two PNe.  More accurate energies for the Se$^{2+}$ $^3$P fine-structure levels are clearly needed.

Forbidden line transition probabilities have been determined for the ground configurations of most detected \emph{n}-capture element ions.  These were computed by Bi\'{e}mont \& Hansen for ions with $4p^n$ valence shells \cite{bh86a, bh86b, bh87, bch88} and by Bi\'{e}mont et al.\ for $5p^n$ valence-shell ions \cite{biemont95}.  In contrast, effective collision strengths for electron-impact excitation have been calculated only for Kr$^{2+}$--Kr$^{4+}$ \cite{schoning97}; Xe$^{2+}$, Xe$^{3+}$, Xe$^{5+}$, Ba$^+$, and Ba$^{3+}$ \cite{sb98}; and Se$^{3+}$ and Zr$^{3+}$.\footnote{K.\ Butler, unpublished}  To our knowledge, effective collision strengths have not been computed for other \emph{n}-capture element ions, and are particularly needed for Br and Rb ions.

In a typical ionized nebula, only one or two ions of a given \emph{n}-capture element can be detected (due in part to the low cosmic abundances of these species, but also to the electronic structure of each ion and the accessibility of strong emission lines).  The conversion of ionic abundances to elemental abundances therefore requires ionization correction factors (ICFs) for the abundances of unobserved ions.  The most robust method of determining the relevant ICFs is through numerical simulations of a nebula's thermal and ionization structure.  However, such models rely on the availability of accurate data for atomic processes affecting the ionization equilibrium of these elements.  For photoionized nebulae such as PNe, these data include photoionization (PI) cross sections and rate coefficients for radiative recombination (RR), (low-temperature) dielectronic recombination (DR), and charge transfer (CT) reactions.\footnote{At photoionized nebula temperatures, processes such as collisional ionization and high-temperature DR negligibly affect the ionization equilibrium.}  Until recent work by our group, these data were not known for the vast majority of trans-iron element ions, thereby rendering \emph{n}-capture element abundance determinations in ionized nebulae uncertain by an estimated factor of two or more \cite{sterling07}.

In the remainder of this paper, we concentrate on recent work that will provide the atomic data needed to determine ICFs and hence elemental abundances of Se, Kr, and Xe -- the three most widely detected \emph{n}-capture elements in PNe and other ionized nebulae.  We have utilized theoretical methods to calculate PI cross sections and rate coefficients for RR, DR, and CT rate coefficients, and benchmarked our results with experimental absolute PI cross section measurements.  We discuss our theoretical work in \S2 and the experimental measurements in \S3, before providing concluding remarks in \S4.

\section{Theoretical Atomic Data Results}

We have computed PI cross sections and RR, DR, and CT rate coefficients for ions of Se and Kr, and similar calculations for Xe are in progress.  While these data are in principle needed for each of the $>$~30 ions of each element, in practice we are able to simplify the problem to the neutrals and first five ions, since PN central stars are not hot enough (generally $<2\times10^5$~K \cite{nap99, stang02}) to significantly ionize species with ionization potentials greater than 100~eV.

For PI, RR, and DR, we have used the atomic structure code AUTOSTRUCTURE \cite{badnell86}, which can efficiently compute the electronic structure, radiative and autoionization rates, and multi-configuration distorted-wave PI cross sections for each ion.  Relativistic effects are accounted for via Breit-Pauli formalism and semi-relativistic radial wavefunctions.  RR rate coefficients were determined from the direct PI cross sections using detailed balance, and DR rate coefficients from the resonant portion of the PI cross sections.  The independent processes approximation \cite{pindzola92} was used to treat RR and DR separately.  The results presented in this section are preliminary, and will be fully detailed elsewhere.

The electronic structure was computed with Thomas-Fermi-Dirac-Amaldi model potentials, and the average of the LS term energies were optimized by varying the radial scaling parameters.  The configuration interaction (CI) expansions and term energy optimizations were adjusted to best reproduce the experimental energy levels and ionization potentials cited in NIST \cite{NIST}.

We computed PI cross sections out of the valence shell for all levels in the ground configuration of each ion, using the CI expansions from the structure calculations.  Ground state PI cross sections for Se and Kr ions near their ionization thresholds are presented in Figure~1.  RR rate coefficients (total and partial final-state resolved) were then computed over the temperature range ($10^1$--$10^7$)$z^2$~K, where $z$ is the charge.  For RR, the CI expansion of the $N$+1-electron ion was expanded to include all one-electron additions to the target configurations.  Total and partial final-state resolved DR rate coefficients were computed over the same temperature range.  For this process, we focused on $\Delta n=0$ core excitations, since high-temperature DR is expected to be negligible at photoionized temperatures.  For DR, the target CI expansions were augmented in order to account for all relevant core excitations, and 1-electron additions to these configurations were included in the $N$+1-electron ion CI expansions.  Following Badnell \cite{badnell06}, we used experimentally-determined target energies whenever possible for our DR calculations, using the theoretical level splittings in cases where the NIST energies are incomplete.

\begin{figure}[ht!]
\centering
\includegraphics[width=12cm,height=16cm]{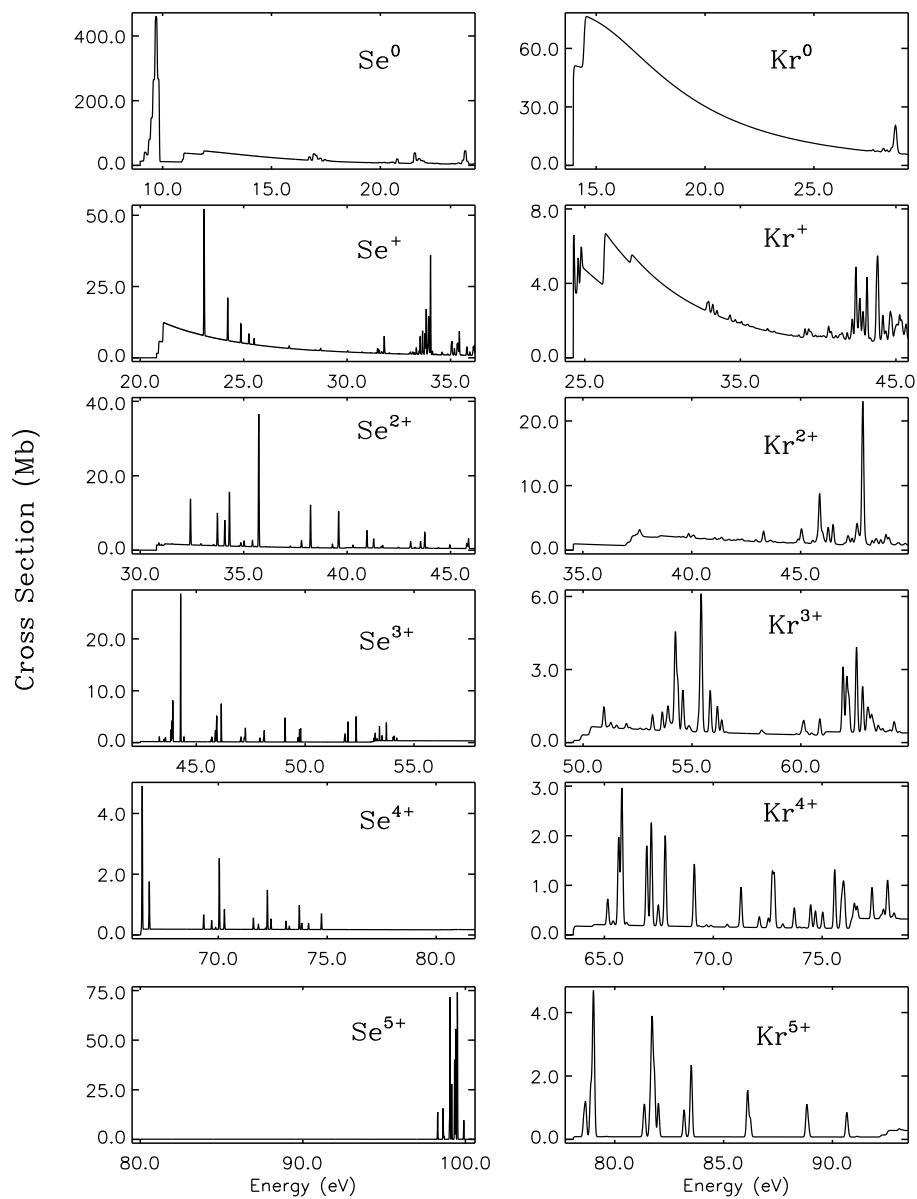}
\caption{Calculated PI cross sections for the ground state of each Se and Kr ion considered, in the energy region near the ground state ionization threshold.}
\end{figure}

We estimated the uncertainties of our results, as these can have significant effects on nebular abundance determinations.  We used three CI expansions of different sizes for each ion, and also tested the sensitivity of our results to other internal parameters of the code (e.g., scaling parameter values for continuum orbitals, the use of non-orthogonal orbitals, the internal interpolation scheme, etc.).  We also compared our results to experimental PI cross sections (see \S 3).

For PI, we estimate typical uncertainties of 30--50\% in the direct cross sections near the ground state ionization threshold.  The largest uncertainties come from using different CI expansions, and are largest for low-charge species.  Similar uncertainties are found for our RR rate coefficients.

Uncertainties in low-$T$ DR rate coefficients primarily arise from the poorly known energies of low-lying autoionizing levels, as only these states are available as DR channels at the temperatures of photoionized nebulae.  The near-threshold resonance positions are notoriously difficult to calculate precisely, leading to large uncertainties in the low-$T$ DR rate coefficients.  Unfortunately,  the energies of these near-threshold states have not been experimentally determined for \emph{n}-capture element ions (or, indeed, for most row~3 and heavier elements \cite{savin99, ferland03}), and hence it is unknown whether they lie just above the threshold (and are available as DR channels) or just below, in which case they do not contribute to DR (though our use of observed target energies alleviates this effect to some extent).\footnote{Recent work \cite{robicheaux10} suggests that the situation may not be as dire as described here.  In finite density plasmas, the continuum threshold is broadened due to thermal considerations, and resonances just below the canonical ionization threshold can contribute to DR.  Thus, the associated uncertainties in the rate coefficient may be decreased.}  To test the sensitivity of our rate coefficients to this effect, we shifted the continuum relative to the near-threshold resonances by an amount equal to the largest energy difference in theoretical vs.\ experimental energies.  These tests revealed uncertainties in our rate coefficients near $10^4$~K (the typical electron temperature of PNe) ranging from a factor of 2 to an order of magnitude for most Se and Kr ions.

At $10^4$~K the DR rate coefficient is greater than that of RR by 1--2 orders of magnitude for low-charge Se and Kr ions, and hence low-$T$ DR is the dominant recombination mechanism.  In photoionization codes used to model ionized nebulae, the large uncertainties in low-$T$ DR rate coefficients necessitates that this process be treated approximately.  As an example (by no means an egregious one!), Cloudy \cite{ferland98} assumes that the low-$T$ DR rate coefficients for the first four ions of Row~3 and heavier elements are the same as the average rate coefficients for C, N, and O ions of the same charge.  However, given the complexity of the target states for near-neutral \emph{n}-capture element ions, there is a much denser set of near-threshold resonances that can contribute to low-$T$ DR than for L-shell systems.  Thus, as atomic number increases the approximation in Cloudy becomes increasingly poor, and for Se and Kr the computed low-$T$ DR rate coefficients exceed the averaged CNO rate coefficients by 1--2 orders of magnitude.  The accuracy of low-$T$ DR rate coefficients is a challenge common to all photoionization codes, and is important to our understanding of the ionization balance of astrophysical nebulae.

CT requires a different treatment, as it is a quasi-molecular problem.  We consider collisions with H atoms of the form X$^{q+}$ + H $\rightarrow$ X$^{(q-1)+}$ + H$^+$ + $\delta$E, where $q=2$--4, since this process is not expected to play a major role in the ionization equilibrium of more highly-charged ions due to the minimal mixing of neutral H with those species in PNe.  We are using a multichannel Landau-Zener code based on the formalism of Butler \& Dalgarno \cite{bd80} and Janev et al.\ \cite{janev83} to determine total and partial final-state resolved CT rate coefficients for Se, Kr, and Xe atoms.  This method has been shown to be accurate to within a factor of $\sim$3 for lighter elements \cite{bd80, kf96}.  Singly-charged ions, however, cannot be treated in this approximation as there are no avoided crossings in the adiabatic potentials (CT is a Demkov process for those species) \cite{stancil01}.  For these systems, we will use a modification of the approximate procedure of P\'{e}quignot \& Aldrovandi \cite{pa86} to estimate the CT rate coefficient.

\section{Experimental Results}

As is true of any atomic data calculation, it is important to benchmark the results to experimental measurements.  To this end, we have measured absolute PI cross sections of Se and Xe ions at the Advanced Light Source (ALS) synchrotron radiation facility at the Lawrence Berkeley National Laboratory in California.  PI cross sections for Kr \cite{lu06a, lu06b, lu_thesis} and Xe$^{3+}$--Xe$^{6+}$ \cite{emmons05, bizau06} had previously been measured, and hence we focused our efforts on the first five Se ions and the first two of Xe.

These measurements were performed using the merged-beams technique \cite{lyon86} at the Ion Photon Beamline (IPB) apparatus \cite{covington02, aguilar03} located on beamline 10.0.1 of the ALS.  In these measurements, a plasma of atomic ions is created within an electron-cyclotron-resonance source and is accelerated with a potential (typically 6~kV).  The ion of interest is selected with an analyzing magnet to form the ion beam, which is collimated with two sets of vertical and horizontal slits and focused with three electrostatic einzel lenses.  A set of spherical electrostatic bending plates merges the ion beam with the photon beam, and along the merged beam path photoionization occurs.  Photoions are redirected to a detector by a demerging magnet and counted.  The photons are produced by electrons in the 1.9~GeV storage ring of the ALS that are accelerated by a 10-cm period undulator.  Photon energies are selected with a spherical monochromator with three adjustable gratings to cover energies between 18 and 300~eV.  The energy resolution is controlled by adjusting the entrance and exit slits of the monochromator.

By measuring the photoion yield as a function of photon energy, the photoionization spectrum is determined in arbitrary units.  To place this on an absolute scale, the photoions must be produced in a well-defined volume.  This is achieved by placing a potential on the ``interaction region,'' a stainless-steel mesh cylinder with a precisely determined length.  The potential energy-tags photoions produced in that region, and the demerging magnet setting is altered to direct only energy-tagged photoions to the detector.  The volume in which the photoions are produced is determined by measuring the overlap of the photon and ion beams with three translating-slit scanners in the interaction region.  Absolute measurements are performed at discrete photon energies to normalize the photoionization spectrum.

We have measured absolute PI cross sections near the ground state ionization thresholds of Se$^+$ (Figure~2, left panel),\footnote{Results have been submitted to J.\ Phys.\ B: At.\ Mol.\ Opt.\ Phys.\ for publication.} Se$^{2+}$, Se$^{3+}$ (Figure~2, right panel)\footnote{Esteves et al., in preparation}, Se$^{5+}$, Xe$^+$, and Xe$^{2+}$.  The Se$^{4+}$ cross section could not be measured due to a large background also seen for the isoelectronic Kr$^{6+}$ (caused by the presence of ions in autoionizing metastable states in the primary ion beam).  Measurement and identification of the resonance structure near the ground state threshold is valuable, since these resonances correspond to low-$T$ DR resonances and can be used to constrain DR rate coefficient calculations.

\begin{figure}[t!]
\includegraphics[width=8cm,height=8cm]{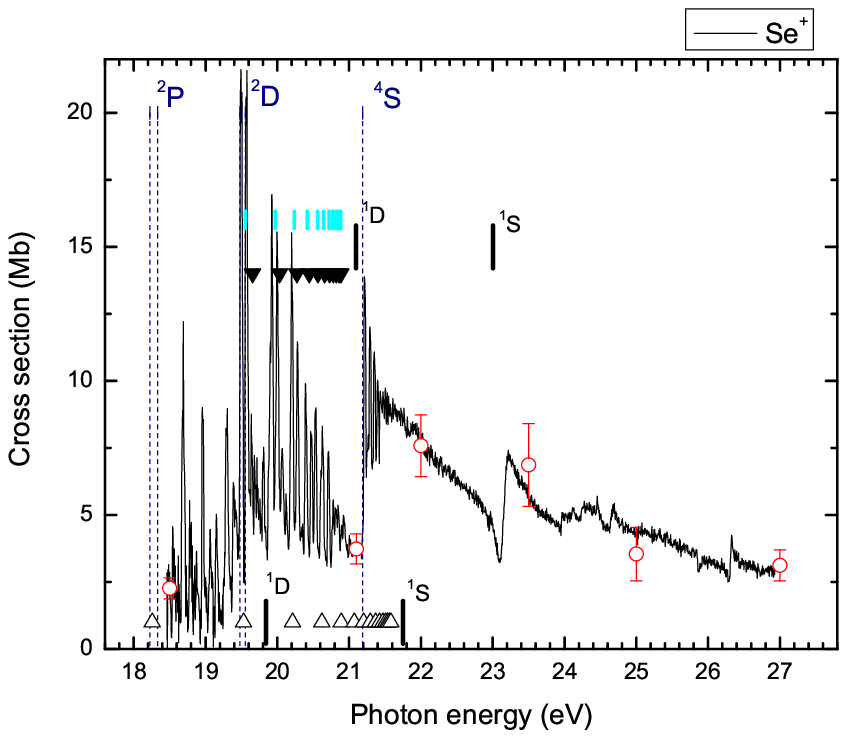}
\hfill
\includegraphics[width=8cm,height=8cm]{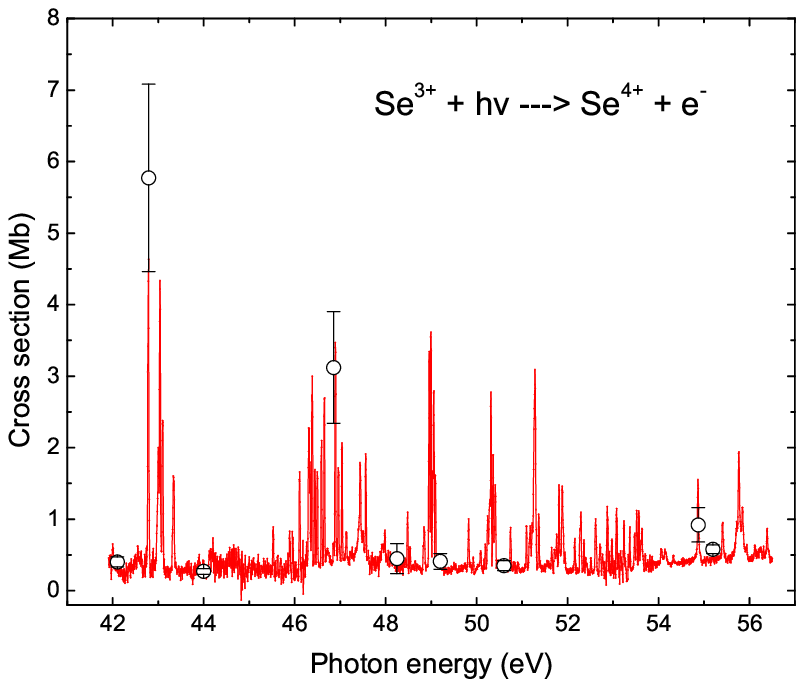}
\caption{\textit{Left panel:} The Se$^+$ absolute photoionization cross section measured at the ALS.  The solid line is the photoionization spectrum normalized to absolute cross-section measurements at discrete photon energies (circles with error bars corresponding to the 90\% confidence level).  Resonances and Rydberg series are indicated.  \textit{Right panel:} The Se$^{3+}$ absolute photoionization cross section, with absolute measurements indicated by circles with error bars indicating the 90\% confidence level.  This figure is modified from \cite{sterling09}.}
\end{figure}

Most atomic ions in photoionized nebulae are in the ground state, and thus PI out of the ground state is of most interest for astrophysical applications.  However, the absolute PI measurements at the ALS are complicated by the fact that the primary ion beam is not composed solely of ground state ions, but contains significant fractions of metastable states with lifetimes exceeding the time of flight through the IPB apparatus.  The metastable fractions can be constrained by measuring the PI cross section below the ground state threshold down to the threshold of the highest energy metastable state (for example, the ground state ionization threshold of Se$^+$ is 21.2~eV; in Figure~2 the cross section and resonances below that energy are all from metastable states), and with the aid of theoretical calculations.

\section{Concluding Remarks}

We have presented preliminary results of an investigation to determine atomic data needed for accurate abundance determinations of the three most widely-detected \emph{n}-capture elements in ionized astrophysical nebulae -- Se, Kr, and Xe.  We are utilizing the atomic structure code AUTOSTRUCTURE \cite{badnell86} to calculate PI cross sections and RR and DR rate coefficients for the first five ions of these elements, and are using a MCLZ code to determine CT rate coefficients.  To calibrate our theoretical results, we have measured the absolute PI cross sections of Se and Xe ions at the ALS synchrotron radiation facility.  We also make use of existing Kr and Xe absolute PI measurements \cite{lu06a, lu06b, lu_thesis, bizau06, emmons05}.

The end goal of determining these new atomic data is to utilize them to derive analytical ionization correction factors (ICFs) for the abundances of unobserved Se, Kr, and Xe ions in ionized nebulae.  Following the methodology of \cite{sterling07}, we plan to incorporate the new atomic data into the photoionization code Cloudy \cite{ferland98}, which we will use to run a grid of models over a range of stellar and nebular parameters (e.g., stellar effective temperature and luminosity, and nebular density) typically encountered in PNe.  ICFs can be derived by finding correlations between the fractional abundances of the observed Se, Kr, and Xe ions and those of routinely detected species such as O or Ar ions, and fitting the correlations with analytical functions.  The new ICFs will be applied to existing spectra of \emph{n}-capture elements in PNe \cite{sharpee07, sterling07, sterling08, sterling09} to reassess the abundances of these three elements.  This will permit robust abundance determinations of \emph{n}-capture elements in PNe (and other types of photoionized nebulae), enabling investigations of \emph{s}-process nucleosynthesis in their progenitor stars at an unprecedented accuracy.

We will also test the sensitivity of abundance determinations to our estimated atomic data uncertainties using Monte~Carlo simulations.  This will help to assess which systems and atomic processes require further analysis.  Additionally, this will highlight the general dependence of abundance determinations on the underlying atomic data that forms the foundation of astrophysical spectroscopy.

\section*{Acknowledgments}

We acknowledge support by the Director, Office of Science, Office of Basic Energy Sciences, of the U.S.\ Department of Energy under contracts DE-AC02-05CH11231, DE-AC03-76SF-00098, and grant DE-FG02-03ER15424. N.\ C.\ Sterling acknowledges support from an NSF Astronomy and Astrophysics Postdoctoral Fellowship under award AST-0901432 and from NASA grant 06-APRA206-0049.  D.\ Esteves acknowledges support from grant NNX08AJ96G with NASA Goddard Space Flight Center and the Doctoral Fellowship Program at the Advanced Light Source.  We thank N.\ Badnell for many enlightening discussions regarding AUTOSTRUCTURE, and P.\ Stancil for helpful discussions and providing a code for our CT calculations.

\bibliographystyle{unsrt_notitle}

\bibliography{asos.bib}

\end{document}